# USING A TERRESTRIAL LASER SCANNER TO CHARACTERIZE VEGETATION-INDUCED FLOW RESISTANCE IN A CONTROLLED CHANNEL

FABRICE VINATIER[1], JEAN-STEPHANE BAILLY[2], GILLES BELAUD [3] & DAVID COMBEMALE[1]

[1] INRA, UMR LISAH, Montpellier, F-34060 France,
e-mail fabrice.vinatier@supagro.inra.fr, david.combemale@supagro.inra.fr

[2] AgroParisTech, UMR LISAH, Montpellier, F-34060 FRANCE
e-mail bailly@agroparistech.fr

[3] Montpellier SupAgro, UMR GEAU, Montpellier, F-34093 France
e-mail gilles.belaud@supagro.inra.fr

**ABSTRACT**

Vegetation characteristics providing spatial heterogeneity at the channel reach scale can produce complex flow patterns and the relationship between plant patterns morphology and flow resistance is still an open question (Nepf 2012). Unlike experiments in laboratory, measuring the vegetation characteristics related to flow resistance on open channel in situ is difficult. Thanks to its high resolution and light weight, scanner lasers allow now to collect in situ 3D vegetation characteristics. In this study we used a 1064 nm usual Terrestrial Laser Scanner (TLS) located 5 meters at nadir above a 8 meters long equipped channel in order to both i) characterize the vegetation structure heterogeneity within the channel form a single scan (blockage factor, canopy height) and ii) to measure the 2D water level all over the channel during steady flow within a few seconds scan. This latter measuring system was possible thanks to an additive dispersive product sprinkled at the water surface. Vegetation characteristics and water surfaces during steady flows from 6 different plant spatial design on channel bottom for 4 plant species were thus measured. Vegetation blockage factors at channel scale were estimated from TLS points clouds and analyzed.

*Keywords*: *3D analysis, LIDAR, blockage factor, experimental channel, vegetation species*

## 1. INTRODUCTION

There is an urgent need for transdisciplinary researches at the interface between ecology and hydrology (Rodriguez-Iturbe 2000). Ditches surrounding vineyards in Mediterranean areas are interesting case studies because both biotic and abiotic matters circulate through these landscape elements. Biotic matters gather a high diversity of fauna and flora, such as weeds and micro-mammals that consider ditches as potential habitats for dispersing. Abiotic matters are constituted of fresh water and sediments that circulate during pluvial episodes. Ecology and hydrology of ditches have been studied separately because of the relative partitioning between disciplines despite of the numerous interrelations existing between biotic and abiotic matters. Water is a potential physical driver for hydrophilic plants and plant architecture has an impact on water flow, for example (Wiens 2002). The partitioning between ecological and hydrological disciplines is stressed by the lack of empirical data on the relationship between plant distribution and water flow, despite some recent studies on this subject (Nepf 2012). Thanks to the recent spread of Terrestrial Laser Scanner (TLS) use in many domains including forestry (Fan et al 2014) and hydraulics (Blenkinsopp et al 2012), we proposed to apply this technique to test the effect of plant architecture on water flow. A complete factorial design with plant species, densities and spatial arrangement were constructed for this purpose in a controlled channel to test the effect of vegetation structure on flow resistance parameter. The objectives of the study is to characterize i) the blockage factor of plants, i.e. the channel cross-section surface ratio occupied by plant material at different water height and flow and ii) to measure the surface water topography during steady flow accordingly, in order to derive further flow resistance parameters at channel scale for hydraulic models.

## 2. MATERIAL AND METHODS

2.1 Material

Four plant species were selected according to their hydrophilic behaviour and branching complexity: *Asparagus acutifolius*, *Scirpoides holoschoenus*, *Elytrigia repens*, and *Lythrum portula*. Approximately 1600 plants of each species were collected in the Languedoc area (43,67N, 3.80W) and kept in plastic boxes filled with fresh water to prevent dessication during the transport. Then each plant were cut at the basis to get calibrated 40 cm's length replicates.

Experiments were conducted in a channel with cement borders that measure 13 meters length, 0.7 meters width and 0.4 meter depth, located in Montpellier, Supagro (43.62N, 3.85W). The channel was chosen according to its dimensions closed to agricultural ditches of the Languedoc vineyards area. The flow of water circulating through the channel was controlled from 25.3 to 51.4 L.s-1 using baffle sluice gates located upstream of the channel. Bottom of the channel was covered with high-density polystyrene foam plates glued using silicone sealant to the ground of the channel. Plates were pierced at a density of 328 holes.m-2 arranged on a regular grid. Holes were filled with plastic dowels to facilitate plants push down and manipulation. Two 30 cm slides were also glued at upstream and downstram of the channel to measure water depth.

Data clouds were measured using a 1064 nm Terrestrial Laser Scanner (TLS) branded RIEGL(R) VZ400 located at nadir above the channel with vertical scan position. For that purpose, the TLS was hanged on the top of a 3 meters' length Hague K12 crane located on a floor at 8 meters up from the channel. Sulphur powder was sprinkled upstream at the water surface during each scan duration (few seconds). Powder composition was chosen according to its high dispersive abilities at water surface and its reflective range in near infrared at the TLS differing from the plants.

2.2 Realization of an experimental design

We tested the effect of plant species, planting patterns and density on the steady flow regime of the channel through four plant species described above for six different planting patterns (regularly planted in staggered rows at a 41, 82, and 164 plants.m-2 (denoted M1, M2, and M3), and according to three different patterns at a 82 plants.m-2 density, denoted M4, M5, and M6), leading to 24 combinations in a complete sampling design. We followed a four-steps procedure for each combination of plant x pattern to keep the same initial conditions at the beginning of the run (Figure 1). Three scans were realized during a combination.

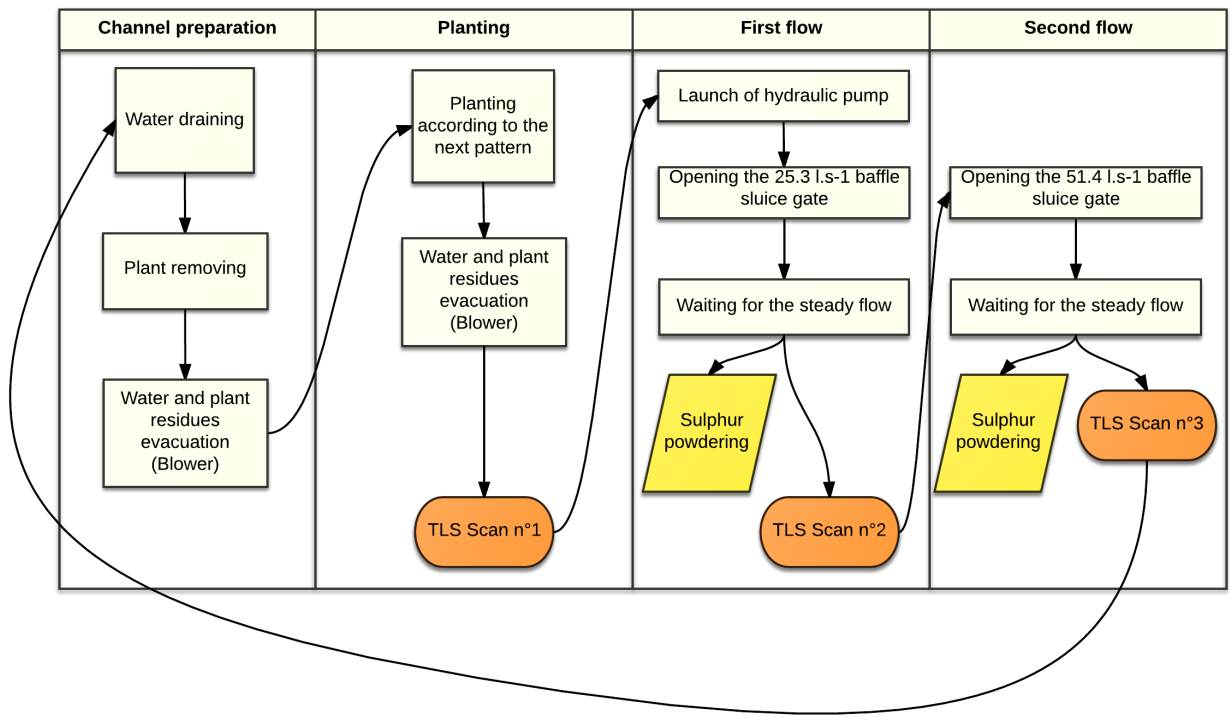

Figure 1: Overview of the steps followed during a single experiment.

2.3 Analysis of the results

TLS scans were first exported in XYZ point clouds using Riscan Pro(R) software. Pulse amplitude of the signal was added for each point. Data clouds issued from the 72 scans were scaled in the same local projection system to get length and width of the channel along X and Y-coordinates, respectively.

Point clouds were analyzed using the R software base and more specific packages (raster, spatstat and rgl). First, blockage factor at local scale was calculated for each cloud considering points number in each voxel of a 3D grid with a given resolution on the clouds issued from the scan without water. Then clouds issued from the water level during steady flow was differentiated from the emerging plants residuals using three filters, applied successively on the cloud: threshold filter on the basis of pulse intensity, nearest-neighbors distance filter (distance from each point to its direct neighbor) and low quantile decimation.

**3. RESULTS**

3.1 Blockage factor estimation at channel scale

Blockage factor corresponds simply to the channel cross-section surface ratio occupied by plant material. It can be defined locally along the channel, at a given cross-section location, or globally at the channel scale by summation for given cross-section width-height location. In this latter case, the mean blockage factor at channel scale does not correspond to a binary image. It can be estimated as a probability at each cross-section width and height location. Thanks to the high density of TLS point clouds and assuming that blockage factor image in channel cross-section is not affected by flow height, we chose to first estimate it by a simple counting process in channel voxels. The resolution chosen for blockage factor calculus was 1 cm3. Blockage factor were summed for each vertical slice of 1 cm width along the channel length divided by the total number of voxels to have a probability between 0 (no blockage) to 1 (complete blockage). They differed greatly between planting patterns and reproduce the vertical plant patterns as expected (Figure 2).

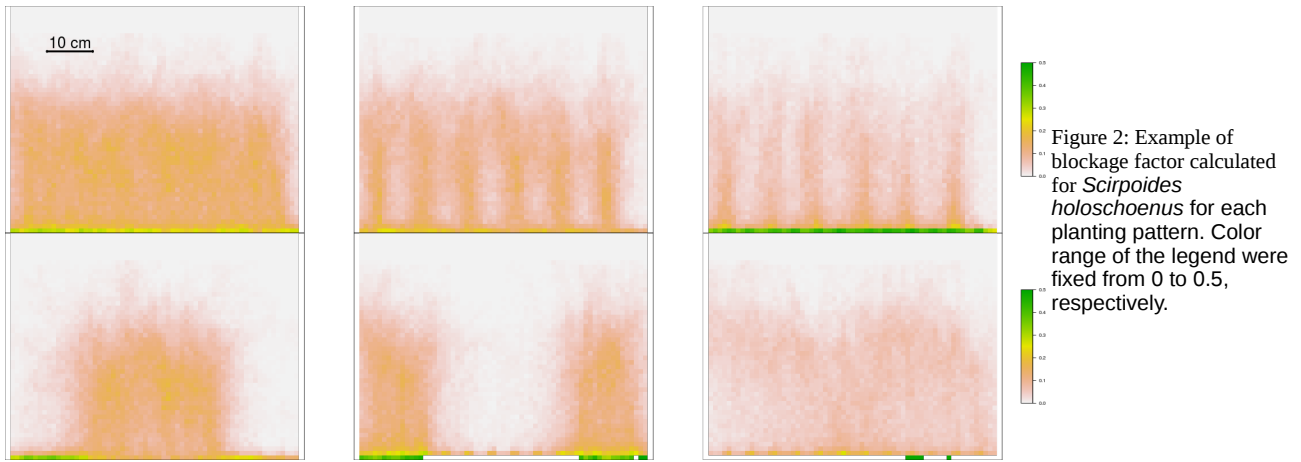

Figure 2: Example of blockage factor calculated for *Scirpoides holoschoenus* for each planting pattern. Color range of the legend were fixed from 0 to 0.5, respectively.

Considering each species individually (Figure 3), lignified plants such as *Asparagus acutifolius* and *Lythrum portula* had the highest blockage factor. Herbaceous plants such as *Elytrigia repens* presented a vertical heterogeneity due to the presence of several blades from each plant sucker laying on the ground, on the contrary to *Scirpoides holoschoenus* that exhibited a higher rigidity level. Despite presenting occlusions for the lignified species due to a single scan position, TLS points clouds gave consistent images of expected blockage factor for each specie.

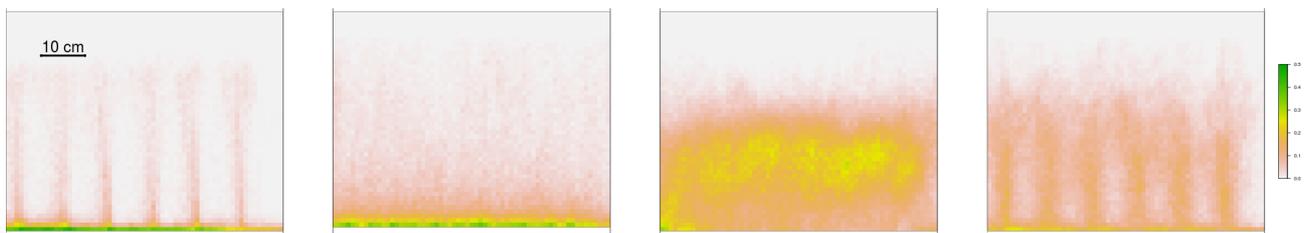

Figure 3: Example of blockage factor estimated for each species for the planting pattern M2 (from left to right: *Scirpoides holoschoenus, Elytrigia repens, Asparagus acutifolius* and *Lythrum portula)*. Color range of the legend were fixed from 0 to 0.5, respectively.

3.2  Water level during steady flow

TLS point clouds acquired during a steady flow at water surface was perturbed by emergent plant branches and leaves. Except outliers coming form multiscattering of laser pulses, no point were theoretically acquired beyond the water surface. In order to get a proper image of the water surface, a process to keep points belonging to the surface was applied as follows: pulse amplitude for classification of points belonging to sulphur powder and plant components (leaves and stem) was defined at 650 DN (Digital Number). Points isolated at a threshold of 5 mm were deleted (nearest-neighbors filter). Decimation of point cloud was realized on the 0.001 quantile. Proportion of points removed compared to the initial cloud using threshold filter, nearest-neighbors filter and quantile decimation was almost 0.86, 0.97, and 0.98, respectively for each cloud (Figure 4).

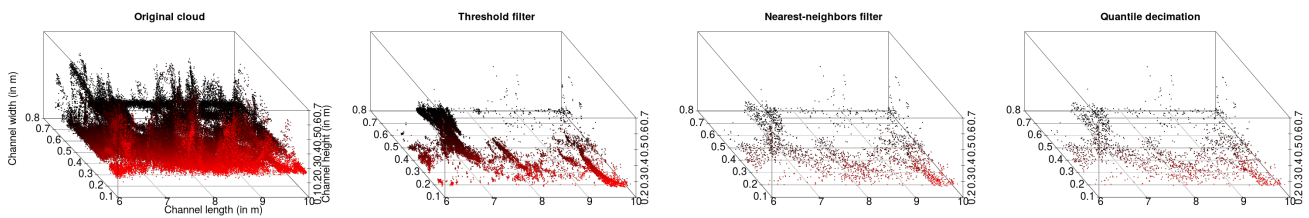

Figure 4: Overview of the filtration steps of the a cloud corresponding to the *Elytrigia repens* species, for planting pattern M3 and first steady flow on a four meters length extract of the channel.

From the remaining points describing the water surface, water level lines (1D) along the channel were first estimated. Water level lines during first and second steady flows, represented in Figure 5, differed greatly between species. Lignified plants such as *Lythrum portula* exhibited the higher water level lines, ranging from 2 cm to 3 cm for first and second steady flows, respectively. Herbaceous plants such as *Scirpoides holoschoenus* exhibited a lower water level line, although significant, from almost 0.5 to 1 cm for first and second steady flows, respectively. Effect of planting density (M1 to M3) was more important than planting pattern (M4 to M6) considering species as varying factor.

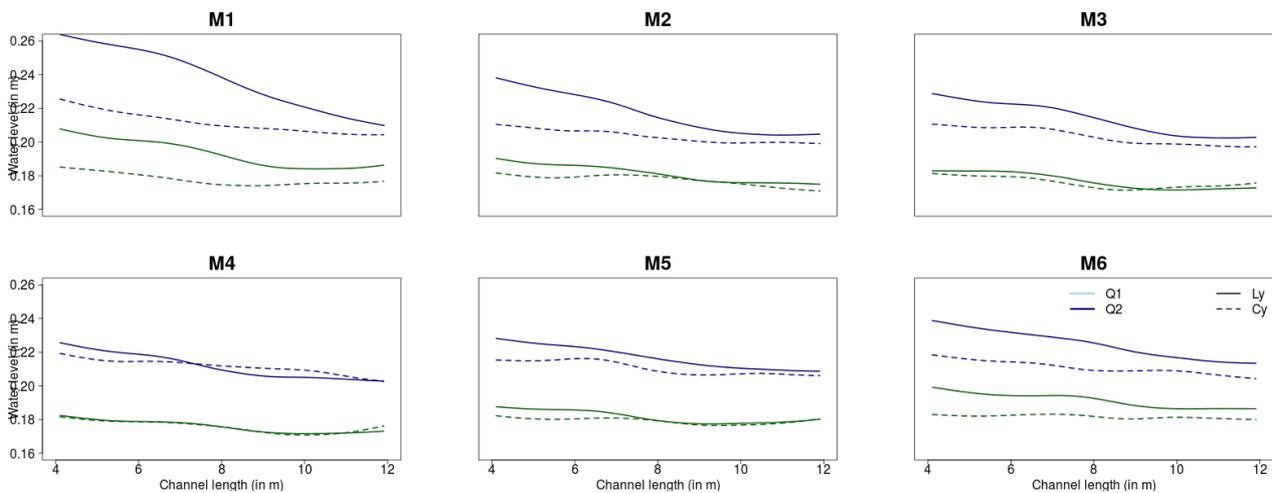

Figure 5: Representation of water level smoothed along the channel for *Scirpoides holoschoenus* (Cy) and *Lythrum portula* (Ly). Green and blue colors correspond to the first (Q1) and second (Q2) steady flow, respectively.

Next we analyzed the mean water level profiles across the channel width, As represented on Figure 6, we demonstrated an evident effect of planting pattern M4 (dense in the middle of the channel, no plant near the sidewalls) and M5 (dense near the sidewalls and no plant in the middle of the channel) on transversal curve of the water level during steady flow. These first consistent results allows the use of TLS to better understand the complex effect of channel bed plant patterns on flows distribution.

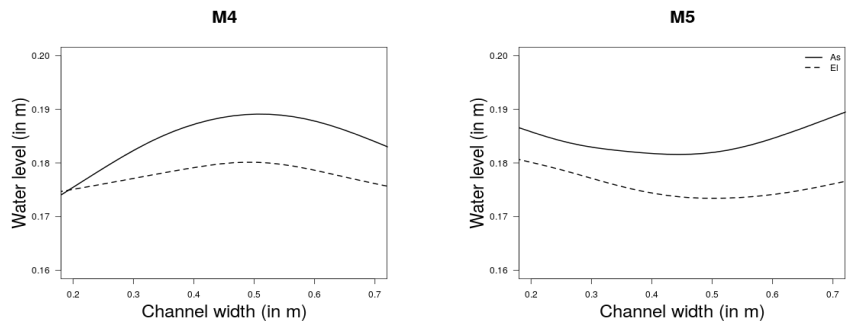

Figure 6: Representation of water level smoothed along the channel length for *Asparagus acutifolius* (As) and *Elytrigia repens* (El).

### 4. CONCLUSIONS

In this paper, the experiments conducted in a equipped channel to disentangle effect of vegetation pattern and density on water discharge showed contrasting results and highlighted the ability of terrestrial lidar scanner to understand the complex link between vegetation blockage factor and water levels within the channels. These first results exhibit a consistent estimation of blockage factor . It also shows the ability of several filtering techniques on TLS point clouds to estimate the effect of planting pattern on both longitudinal and transversal variation of the curvature of water level. These estimation of 2D water levels offers new possibilities to better link immersed vegetation characteristics and synthetic resistance parameters for hydraulic modeling.

**ACKNOWLEDGMENTS** Authors are grateful to the ONEMA and the INRA institutions who both funded the experimental work exposed in this paper within the 'Fossés Infiltrants et Pesticides' project (ONEMA) and the 'Pari-Scientifique: Hydro-écologie des fossés agricoles' project (INRA-EA department).